# Entrainment of Lymphatic Contraction to Oscillatory Flow

Anish Mukherjee[1], Joshua Hooks[2], J. Brandon Dixon[2,3]

[1]Department of Electrical and Computer Engineering, [2]Department of Mechanical Engineering, [3]Department of Biomedical Engineering, Georgia Institute of Technology, Atlanta, GA 30332

## Abstract

Lymphedema, a disfiguring condition characterized by the asymmetrical swelling of the limbs, is suspected to be caused by dysfunctions in the lymphatic system. Lymphangions, the spontaneously contracting units of the lymphatic system, are sensitive to luminal wall shear stress. In this study, the response of the lymphangions to dynamically varying wall shear stress is characterized in isolated rat thoracic ducts in relation to their shear sensitivity. The critical shear stress above which the thoracic duct shows substantial inhibition of contraction was found to be significantly negatively correlated to the diameter of the lymphangion. The entrainment of the lymphangion to an applied oscillatory shear stress was found to be significantly dependent on the difference between the applied frequency and intrinsic frequency of contraction of the lymphangion. The strength of entrainment was also positively correlated to the applied shear stress when this shear was below the critical shear stress. The results suggest an adaptation of the lymphangion contractility to the existing oscillatory mechanical shear stress as a function of its intrinsic contractility and shear sensitivity. These adaptations might be crucial to ensure synchronized contraction of adjacent lymphangions through mechanosensitive means and might help explain the lymphatic dysfunctions that result from impaired mechanosensitivity.

## Introduction

The lymphatic system plays a crucial role in the regulation of tissue fluid balance and hence, the maintenance of interstitial fluid volume. The interstitial fluid that accumulates as a result of extravasation of fluid from the blood capillaries is cleared by the lymphatic system. Through a network of capillaries and collecting vessels, the lymphatic system takes up the excess interstitial fluid and transports it back into the blood circulation. The lymphatic system plays an important role in not only fluid homeostasis but also in lipid transport and the immune system. Dysfunction of the lymphatic system is closely related to a condition called "lymphedema", which is characterized by persistent swelling of the tissue space (often the limbs) due to an accumulation of fluid [1–3]. Lymphedema has largely been understudied since it is not a life-threatening condition, but it has a severe impact on the quality of life of the patient. In addition to a gross swelling of the arms and/or legs, which is aesthetically unpleasant and can lead to psychological scarring for the patient, the freedom of movement of the limbs is restricted leading to discomfort and hindered functionality. In recent years, the severity of lymphedema has been recognized by the research community and focus has been put on studying the underlying mechanisms governing this condition.

The onset and progression of lymphedema is coincident with a change in the contractile function of the lymphatic system, possibly due to aberrant mechanical forces [1–8]. Lymphangions, the basic functional units of the collecting lymphatic system exhibit intrinsic contractility. The responsiveness of lymphangions to mechanical stimuli has been well documented [9–19]. There are primarily two types of mechanical forces that need to be considered in the lymphatic system – the hoop stress exerted on the wall due to the transmural pressure and the wall shear stress in the lumen of the collecting lymphatic vessels due to fluid flow. The effect of luminal pressure on the lymphangion is that of exerting a hoop stress (stretch) on the lymphatic endothelial and muscle cells, and its impact on the physiology of the collecting vessels has been studied widely [14,16,20–22]. The transmural pressure has a direct influence on the contraction frequency, the end systolic and diastolic volume of the vessel, as well as other pumping metrics like ejection fraction. The general trend is an increase in the contraction frequency and pump function in response to an elevated transmural pressure from 0 to some optimal pressure, after which further increases in pressure cause a reduction in pump function [6,16,21,23–27].

Shear stress is an important factor, modulating both the tonic and phasic contraction of the lymphangions [12,15,18,19,21], but has been less widely studied compared to the effect of transmural pressure on lymphatic pump function. Wall shear stress is directly related to the lymph flow rate through the vessel and is also dictated by the vessel diameter. An elevation in lymph flow rate corresponds to an increased wall shear stress and a reduction in the contractility and tone of the vessel results in a lower shear stress. Experimentally, it is easier to control the pressure gradient across the vessel than the flow rate through it and hence studies have mostly focused on using the pressure gradient as a means to impose wall shear stress on the vessel. Gashev et al showed that both the chronotropic and inotropic response of the lymphatic vessels are negatively affected by an imposed pressure gradient [18]. Numerous studies since these original ones have confirmed the findings of Gashev et al, that imposed flow decreases both the contraction amplitude and frequency [12,28–32]. With an increase in the applied pressure gradient in rat mesenteric lymphatic vessels, the contraction frequency decreases, but this decrease is only temporary since the frequency increases over time with the same applied pressure gradient. The contraction amplitude decreases due to an increase in the end-systolic diameter. This effect is more pronounced in rat thoracic ducts [33]. These changes are also dependent on the rate of change of the flow, with a higher rate of change producing a larger change in the lymphatic pumping parameters [12,14].

The region chosen for the study can affect the response of the lymphatic vessels to transmural pressure and wall shear stress. For example, contraction frequency was found to be more heavily dependent on transmural pressure for rat mesenteric and cervical lymphatics as opposed to the thoracic duct [33]. Rat thoracic ducts also significantly inhibit their contraction with large favorable pressure gradients, while rat mesenteric and cervical lymphatics still exhibit contraction at these pressure differences [18,34]. Since these studies controlled the flow through the vessel by controlling the pressure at the inlet and the outlet of the cannulae, they did not take into consideration the exact wall shear stress being applied, or the fact that the relation between the applied pressure difference and the wall shear stress is highly dependent on the diameter of the vessel and the pressure drop that occurs along the pipette tips. Thus it is important to investigate the mechanosensitivity of the lymphangions with respect to the wall shear stress experienced by the lymphangion.

While the response of lymphatic vessels to oscillatory shear stress have been explored before [12], we show that the response is more nuanced than previously believed. The response of lymphatic vessels to oscillatory flow is hypothesized to be a consequence of the inhibition in contraction that is observed in response to elevated shear stresses. Thus the ability of a lymphatic vessel to synchronize its contraction to an externally applied shear stress is hypothesized to be dependent on the frequency and amplitude of the applied shear stress and also on its sensitivity to wall shear stress and intrinsic contraction frequency (the average frequency with which the vessel contracts when no flow is applied to its lumen). A platform is introduced to perform these studies in the frequency domain using continuous wavelet transform. This platform allows for the analysis of the response of the lymphatic vessels to dynamic forces by looking at their frequency domain response, which allows for the quantification of their entrainment to externally applied oscillatory forces.

## Results

### *Characterization of wall shear stress sensitivity*

To determine the wall shear stress sensitivity vessels were exposed to a favorable pressure gradient that was linearly increased in magnitude over a fixed time interval, while holding the transmural pressure constant while video of the contracting vessel was recorded through a microscope. Vessel frequency over time was calculated using a continuous wavelet transform (CWT), which is a frequency domain analysis technique that provides the spectral content of a waveform as a function of frequency and time. The bases for the CWT are the wavelets that are defined by their scale and the shift. The scale of the wavelet can be converted to a frequency by knowing the sampling frequency of the signal and the center frequency of the wavelet. The magnitude of the wavelet coefficients are obtained as a function of time and scale (and hence frequency) and are plotted as a surface. An example spectrogram is shown in Fig. 1a for a diameter tracing of a contracting vessel at a single location along the vessel wall. At any particular point in time, the frequency at which the magnitude of the coefficient (hereby referred to as "power") is the highest was defined as the "dominant frequency" at that time point which is represented in Fig. 1b. The use of CWT allows the subsequent analyses of the response of the lymphangion to different wall shear stress conditions, by studying the spectral composition of the signal as a function of the frequency and/or amplitude of the applied wall shear stress.

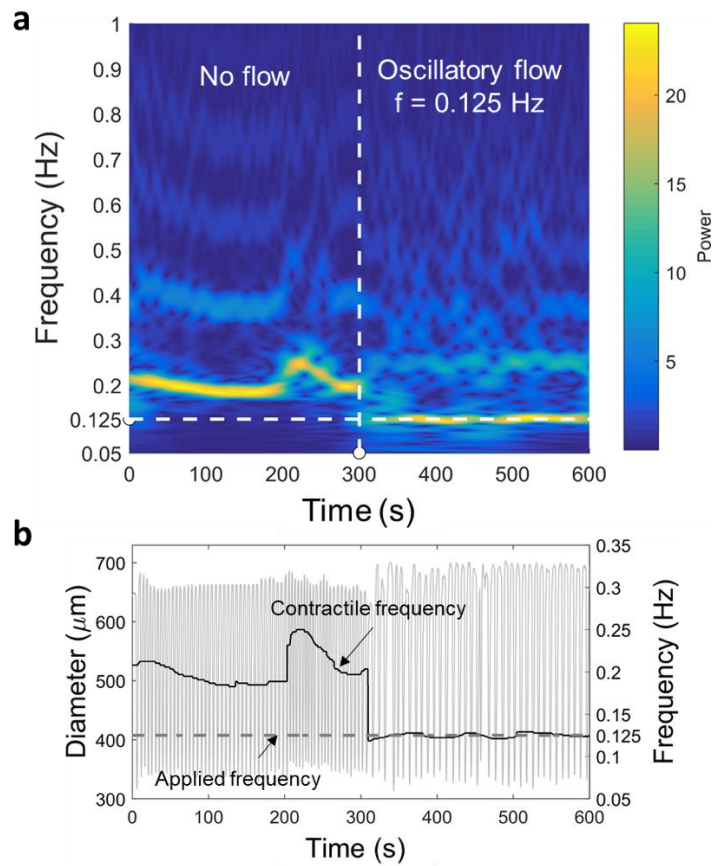

Fig. 1: The method for analyzing the data in the frequency domain is shown. a) CWT provides detailed frequency information. The vertical dashed line represents the time at which imposed oscillatory flow begins and the horizontal dashed line represents the frequency of the imposed flow. b) The frequency information can be easily isolated from the CWT. The calculated frequency is shown as a function of time and is overlaid over the diameter variations. For this particular experiment, the vessel was exposed to no flow at the start of the experiment and at 300 seconds an imposed oscillatory flow was applied to the vessel at the frequency indicated in the figure.

From previous work it is known that the wall shear stress is necessary to inhibit contraction within different thoracic duct segments is quite variable [12]. Hence it is important to establish the wall shear stress (WSS) sensitivity of each vessel prior to exposing them to oscillatory wall shear stress (OWSS) in order to determine the extent that each vessel's unique shear sensitivity is responsible for entrainment to oscillatory flow. To this end, the response of the thoracic duct to a ramped shear stress was analyzed with CWT to obtain the shear sensitivity information. The dominant frequency was obtained as a function of time from the CWT of the diameter tracings for the ramp experiment. The imposed WSS was then obtained as a function of time for the ramp protocol by reading the syringe position of the perfusion system, assuming Poisuielle flow and using the diastolic diameter of the vessel as described previously [12]. A linear fit to this data was obtained, assuming that the imposed WSS is 0 at the start of the shear ramp.

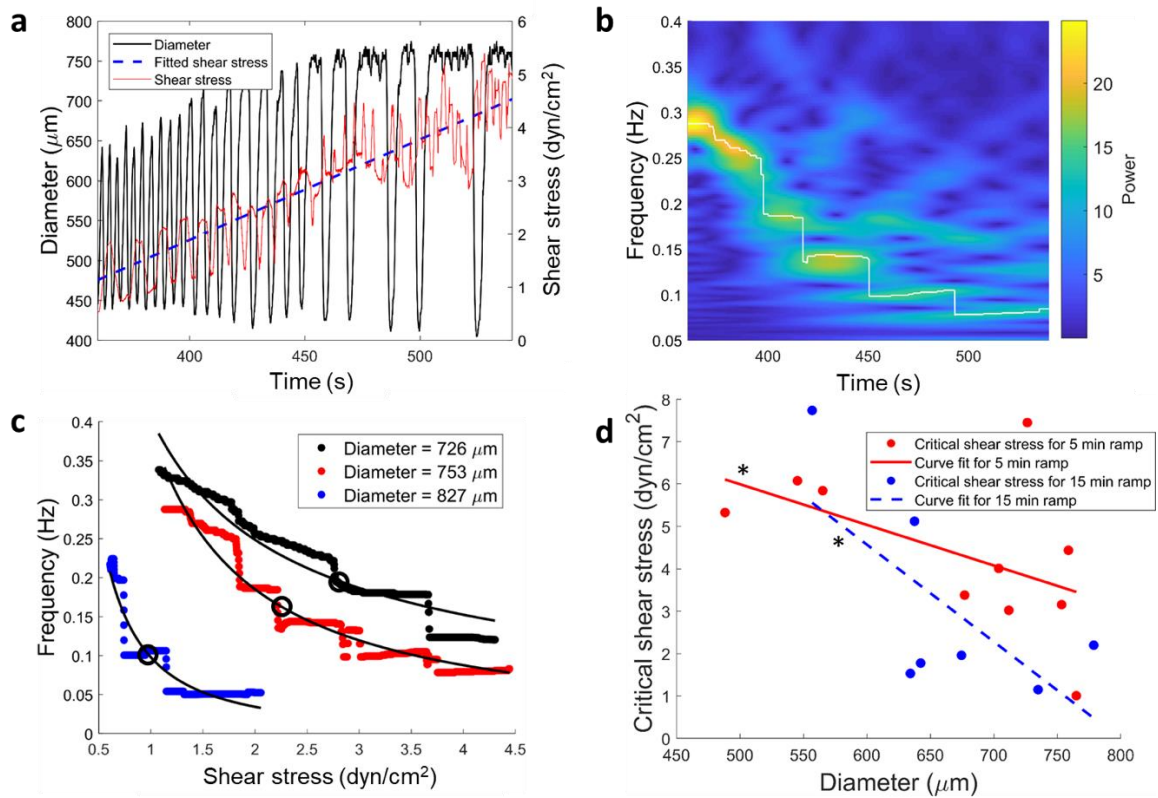

Fig. 2: The critical wall shear stress of the vessel negatively correlates with the passive vessel diameter. a) The contraction frequency of the vessel goes down with time when a ramped shear stress is applied to the vessel. This can be observed from the diameter tracing shown along with the applied shear stress. b) The same information is represented in the CWT spectrogram where the decrease in frequency can be quantified from the dominant frequency information. c) The frequency, when plotted as a function of the shear stress, shows the characteristics of a power law curve. Fitted curves are shown for three different vessels and a dependence on the diameter of the vessel can be readily seen. d) The critical shear stress is plotted as a function of the diameter of the thoracic duct for the two different ramps applied to the vessel. Linear regression curves are fitted to the data and show a significant negative correlation (Spearman correlation coefficient of -0.7454 and -0.787 with p values of 0.0133 and 0.0357 respectively for 5 min and 15 min ramps). The regression does not depend significantly on the rate of the ramp applied (p = 0.8629).

With the fitted shear stress data being obtained, the dominant frequency was then plotted as a function of the shear stress to obtain the shear sensitivity (Fig. 2). The relationship between frequency and WSS appears to follow a power law function of the form $f = at^b$, where f is the frequency, t is the time, and a and b are the parameters to be optimized. Curves of this nature fitted to the data had an average r-squared value of about 0.9, which indicated that the power law model provides a good approximation to the shear-frequency relationship over the ranges tested.

The shear sensitivity of the vessel was represented by a "critical shear stress", which was defined as the shear stress at which the frequency of contraction drops down to half of the intrinsic frequency of contraction of the vessel. The critical shear stress was obtained from the power law relationship relating the frequency of contraction to the imposed shear stress. The shear sensitivity is inversely related to the critical shear stress. The critical shear stress ranges between 0.1 to 10 dynes/cm$^2$ and is found to be significantly correlated to the average diastolic diameter of the vessel, irrespective of the rate of the ramp waveform applied to the vessel. The Spearman correlation coefficient was calculated between the critical shear stress and the diastolic diameter and were found to be -0.7454 and -0.787, with p values of 0.0133 and 0.0357 respectively for the ramp waveforms applied for 5 min and 15 min. The significance of the correlation coefficient was calculated using Fisher's r-to-z transformation. The shear sensitivity was not found to be significantly related to the rate of the ramped shear stress applied to the vessel (p = 0.8434).

*Effect of oscillatory shear stress*

To study the effect of variations in the frequencies and amplitudes of the OWSS on the response of the lymphangion, the spectral distribution of power and how it changes with the frequency and amplitude of the shear stresses applied to the vessel was investigated using CWT. The percentage of power at a particular frequency is representative of the amplitude of contraction of the vessel at that frequency. The percentage of power in the applied frequency is thus indicative of the "strength of entrainment" of the vessel to the applied frequencies. Vessels were exposed to three different frequencies of imposed flow (0.075 Hz, 0.2 Hz, and 0.35 Hz). These values were chosen to be reflective of frequencies that were less than, comparable to, and above the typical intrinsic contraction frequency of isolated pressurized rat thoracic ducts. These three frequencies were also applied with three different amplitudes of the imposed pressure gradient waveform ($\Delta P$ = 4, 6, and 8 cm $H_2O$, where $\Delta P$ is the pressure difference measured between the inlet and the outlet pipettes on which the vessel is cannulated).

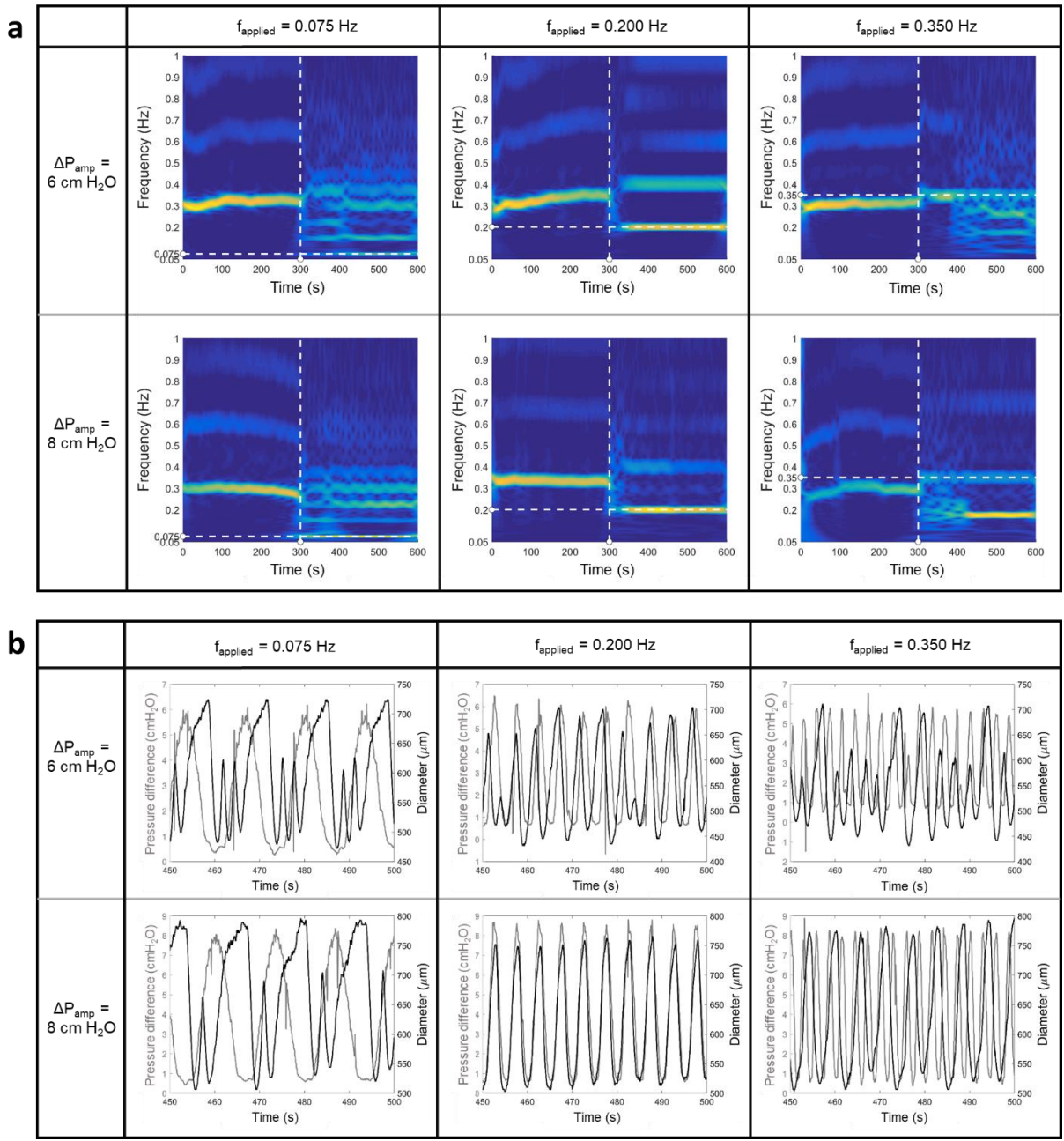

Fig. 3: Representative CWT spectrograms and diameter and pressure gradient tracings are shown for the three applied frequencies of 0.075 Hz, 0.2 Hz and 0.35 Hz and two pressure gradients having amplitudes of 6 cmH$_2$O and 8 cmH$_2$O. a) The graphs show a clear dependence of the entrainment, represented by the power in the applied frequency component, with the applied frequency and the magnitude of the shear. When the applied frequency exceeds the intrinsic frequency of contraction (about 0.3 Hz), the maximum power is seen to be localized at half the applied frequency instead of the applied frequency. b) The dependence of the entrainment on the applied frequency and shear stress is also seen from the pressure and diameter tracings. When the applied frequency is greater than the intrinsic frequency of contraction, the vessel contracts at half the applied frequency.

An observation of the CWT spectrograms for a single vessel at different applied frequencies and amplitudes, as shown in Fig. 3a, reveals that the extent of entrainment of the vessel depends on both the frequency of the applied waveform and the magnitude of the pressure gradient amplitude. An interesting thing to note is that for an applied frequency of 0.35 Hz, the maximum power did not occur at the applied frequency, but rather at half the applied frequency or at 0.175 Hz. This phenomenon is seen for a lot of vessels that have an intrinsic contraction frequency less than 0.35 Hz. This agrees with the hypothesis that the entrainment seen in the vessels is actually an effect of shear inhibition when the WSS rises above the critical shear stress. Since the frequency of oscillation of the WSS is more than the vessel's intrinsic contraction frequency, and since the entrainment is hypothesized to happen as a result of lowering of the intrinsic contraction frequency, it gets entrained at a lower frequency, which is seen to be half the applied frequency. These spectrograms hence show an entrainment of the contraction that is dependent on the applied frequency as well as the intrinsic contraction frequency. Furthermore, the power is higher when the applied pressure gradient is higher, indicating that the entrainment is higher at higher shear stresses.

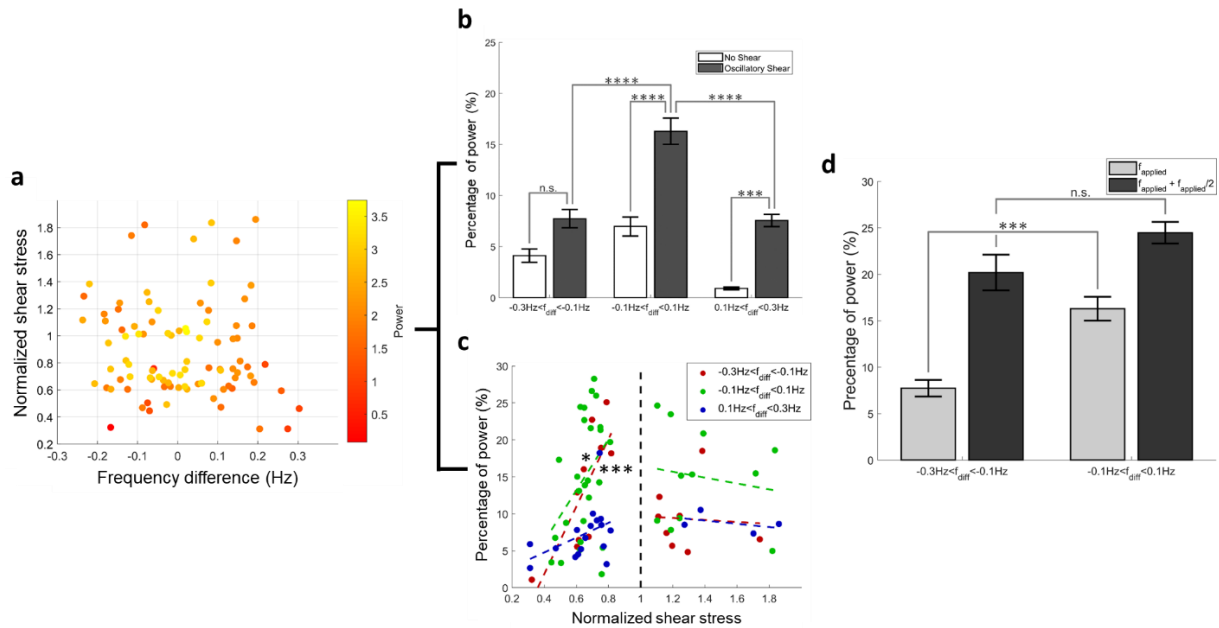

Fig. 4: The entrainment of the vessel depends on both the frequency and magnitude of the imposed oscillatory wall shear stress. a) The percentage of power representing the entrainment is shown w.r.t. both the frequency difference and the shear stress difference. b) The percentage of power is shown at different ranges of frequency under no shear and oscillatory shear conditions, along with the standard error. c) The entrainment is significantly correlated to the normalized shear stress (maximum WSS under imposed oscillatory conditions normalized to the critical WSS for the vessel) when the applied shear stress is below the critical shear stress, for similar and high applied frequencies. d) When the sum of the percentage of power at both the applied frequency and half the applied frequencies are considered, there is no significant difference between the high and medium applied frequencies. This suggests that more power is present at half the applied frequency components when the externally applied frequency is more than the intrinsic frequency of contraction of the vessel.

The entrainment of thoracic ducts (n=8) to the applied OWSS was investigated w.r.t. the difference

Page | 6

between the intrinsic frequency of contraction and the applied frequency, as well as the applied shear stress normalized to the critical shear stress for the vessel (Fig. 4a). The effect of the frequency difference (calculate as the intrinsic frequency minus the applied frequency) on the entrainment of the vessel to the oscillatory flow was investigated by dividing the dataset into 3 windows corresponding to applied frequencies that were higher than (-0.3 Hz to -0.1 Hz), similar to (-0.1 Hz to 0.1 Hz) and lower than (0.1 Hz to 0.3 Hz) the intrinsic frequency as seen in Fig. 4b and 4c. An unbalanced two-way ANOVA test was performed with an alpha level of 0.05 to obtain the statistical significance of the entrainment w.r.t. the frequency difference and the flow conditions. The entrainment of the vessel to the oscillatory shear stress is evident from the significantly increased percentage of power in the applied frequency between no shear and oscillatory shear conditions ($p = 0.0002$, 2.07e-08 and 0.0028 for the three frequency groups). The entrainment was found to be maximum when there is minimal difference between the intrinsic frequency of contraction and the applied frequency. The percentage of power at low applied frequencies was found to be significantly lower than when the applied frequencies were similar to the intrinsic contraction frequency ($p = 3.77e-08$), demonstrating that the entrainment is hampered at low applied frequencies. To investigate the effect of the magnitude of the applied shear stress on the entrainment, the maximum applied shear stress was normalized w.r.t. the critical shear stress for each vessel so as to incorporate the unique mechanosensitivity of each vessel into the analysis. The entrainment was found to be significantly correlated to the shear stress applied to the vessel (using Fisher's r-to-z transformation) in the similar and high frequency bands (r-squared values of 0.1643 and 0.6141 with $p = 0.0399$ and 0.0073 respectively) when the maximum applied shear stress was below the critical shear stress as seen in Fig. 4d. The entrainment was not significantly correlated to the shear stress for vessels where the maximum applied shear stress was above the critical shear stress, suggesting that there is no added benefit to entrainment from a higher WSS when the WSS is already above the critical shear of the vessel. The possible involvement of shear inhibition as a mechanism for lymphatic entrainment is also supported by Fig 4d. The significant difference in the entrainment between the applied waveforms with higher and similar frequency to the intrinsic frequency is lost when the power at both the applied frequency and half the applied frequency are considered. This suggests a higher concentration of power at half the applied frequency when the external flow waveform has a higher frequency than the intrinsic frequency and demonstrates that the vessels are contracting at half the applied frequency when the externally applied frequency exceeds the intrinsic frequency.

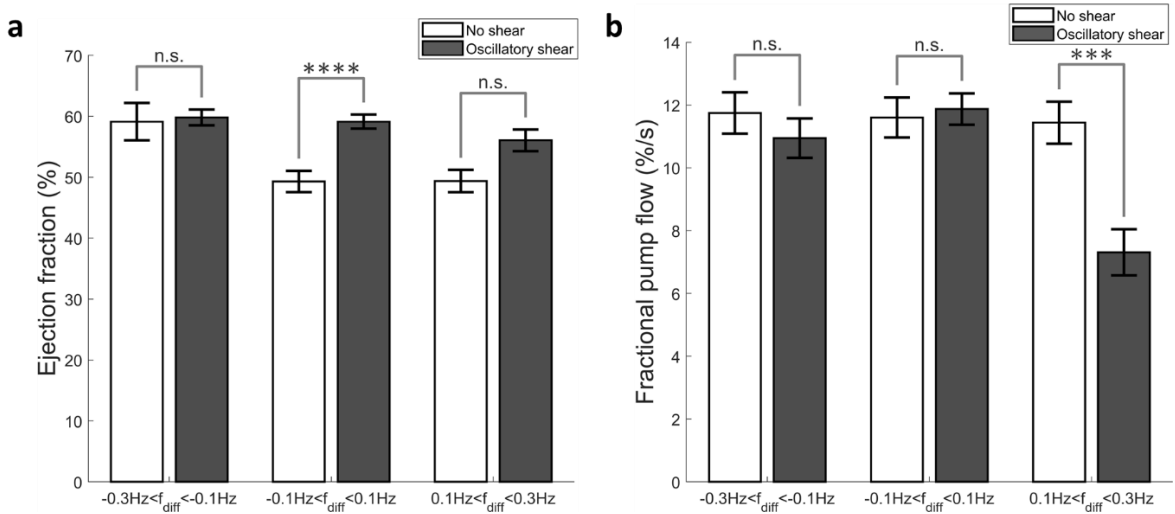

Fig. 5: Ejection fraction and fractional pump flow are differentially modulated by imposed flow. a) Applying an OWSS waveform to a vessel at a frequency similar to the intrinsic contraction frequency enhances ejection fraction ($p = 3.84e-05$). b) The fractional pump flow is significantly lower when low frequency waveforms are applied ($p = 2.8e-04$)

Two pump flow metrics; the ejection fraction which represents the percentage of lymph that is pumped by the lymphangion per contraction, and the fractional pump flow which represents the percentage of lymph that is pumped by the lymphangion per second; were investigated as a function of the frequency of the shear stress applied to the vessel and the flow condition, as seen in Fig. 5. An unbalanced two-way ANOVA test was performed with an alpha level of 0.05 to test for statistical significance of the entrainment w.r.t. the frequency difference and flow conditions. The ejection fraction was found to significantly increase when the applied frequency was close to the intrinsic frequency ($p = 3.84e-05$) as seen in Fig. 5a. The fractional pump flow, representing the fluid pumped as a result of lymphatic contractility (percentage of the fluid volume per second) was found to be significantly lower when the applied frequency was less than the intrinsic frequency, as seen in Fig. 5b ($p = 2.84e-04$ between the no shear and oscillatory shear conditions). This is a consequence of the contraction frequency of the vessel getting lowered as a result of the imposed flow and hence reducing the fractional pump flow.

## Discussion

The lymphatic vessels, owing to their intrinsic contractility are exposed to a unique mechanical microenvironment that sets them apart from the blood vasculature. Unlike blood vessels, the lymphatic vessels undergo both slow, long-term tonic contractions as well as fast, short-term phasic contractions. This unique contractility subjects the lymphangions to oscillatory transmural pressures and wall shear stresses due to the contractility patterns of upstream and downstream lymphangions. During a typical contraction cycle the lymphangion can contract up to 39% of its diameter and while average wall shear stress over the contraction cycle is on the order of 0.6 dyn/cm$^2$, the peak shear stress is around 8 dyn/cm$^2$ as found in rat mesenteric lymphatics[35]. When exposed to edemagenic conditions[36,37], or in response to elevated loads due to high-fat meals[38], the flow rate and thus maximal shear experienced by the vessel can increase 10-fold to about 5 to 40 dynes/cm$^2$ [36,37]. In comparison, the blood vasculature of similar dimensions as the lymphatics, such as small arteries in rats typically show average wall shear stress in the order of 15 dynes/cm$^2$ [39]. Smaller arteries and veins can show even higher average shear stress, going up to 70 dynes/cm$^2$ [40–42]. Hence the lymphatic vessels are subjected to a mechanical microenvironment characterized by a low and oscillatory shear stress, albeit with maximal transient WSS values that can approach that of the blood vasculature.

When exposed to imposed flow ex-vivo (in which the WSS imposed was previously unknown), lymphatic vessels are known to inhibit their contraction [15,18]. This shear dependent inhibition of lymphatic contractility is thought to be an important mechanism in their transition from a pump to a conduit for lymph, depending on the external forces, in order to optimize lymph flow [32]. Shear mediated inhibition of contractility of the lymphangions, similar to the flow mediated dilation seen in blood vasculature, in endothelium dependent and has been reported to be dependent on the region chosen for study [33]. In vitro studies have also shown the involvement of the lymphatic endothelium in this flow mediated response [43,44]. However the actual WSS magnitudes at which this contraction inhibition occurs, has been limited to one report on a small set of vessels [12]. As exemplified by Fig. 2, the frequency of contraction of the vessel is found to be inversely related to the shear stress in the vessel. Under sufficiently high shear stresses, the vessel might stop contracting altogether. A critical shear stress can therefore be defined as an indicator of a threshold for the pump-conduit transitions for the lymphatic vessels. Further, this critical shear stress is inversely related to the diameter of the vessel, indicating that larger vessels are more sensitive to the wall shear stress, which is agreement with observations made in the literature comparing mesenteric lymphatic vessel and thoracic duct [33]. This alludes to an adaptation of the lymphangions to their own unique microenvironment in order to optimize their lymph transport capabilities.

The synchronized contraction of the lymphangions in order to pump lymph has long been observed in vivo and have been suggested as being important for the optimal transport of lymph depending on the load that the vessels are being subjected to [45,46]. The lymphatic vessels seem to have developed so as to utilize the propagation of depolarization waves along the lymphatic muscle layer for their contraction, facilitated by the electrical decoupling between the lymphatic endothelial and muscle cell layers [47]. Further, the role of endothelium derived relaxation factors like nitric oxide and histamine in the flow mediated dilation have been investigated [18,28,29,31], and their presence was found to be correlated to the shear stress applied on the vessels. Computational models have also shown that the presence of this mechanosensitivity would facilitate the coordinated pumping of lymphangions [48,49]. Our previous work showed that the application of oscillatory shear stress can cause entrainment in the contraction of lymphangions, and the entrainment is lost if the shear stress is sufficiently low [12]. Hence the existing literature suggests that the entrainment of contraction between lymphangions is affected at least in part by the shear sensitivity of the lymphangions, thus implicating flow mediated dilation as a direct causative factor in the coordinated contraction of a chain of lymphangions.

The idea that entrainment is caused by flow mediated dilation is supported by the result as represented in Fig. 4 where the percentage of power in the applied frequency, representing the degree of entrainment, was found to be significantly higher when the applied frequency is close to the intrinsic frequency of contraction. This indicates the presence of an optimum frequency of contraction that the lymphatic vessels have developed, possibly as a result of their microenvironment. Since the entrainment is a result of inhibition of contraction by flow, it is not possible for the vessel to contract with a frequency higher than its intrinsic contraction frequency. This is supported from our observation that when a frequency of 0.35 Hz is applied to a vessel having an average intrinsic contraction frequency of about 0.3 Hz or lower, the vessel is unable to contract at the imposed frequency, and thus ends up contracting at half the applied frequency, effectively inhibiting its contraction at every other applied WSS peak. Further supporting a mechanism involving flow-mediated dilation is the observation that the extent of entrainment of the vessel to the imposed flow is highly correlated with the ratio of the maximum imposed wall shear stress and the intrinsic critical shear. The higher the magnitude of the peak shear, the stronger the synchronization. Interestingly, this correlation is lost when comparing vessels in which the peak imposed WSS was above the measured critical shear, suggesting that elevating the WSS stress above this critical shear value has no benefit to further ensuring contractile entrainment.

The potential physiologic benefit of these unique features of vessel entrainment to flow is perhaps best illustrated in Figure 5. Under conditions were the imposed flow rate frequency is close to the intrinsic contractility of the vessel, the presence of this oscillatory flow enhances the ejection fraction of the vessel compared to the ejection fraction that is observed in the vessel before the application of oscillatory flow. In chains of lymphangions one would expect the frequency of the flow waveform resulting from ejection of fluid from the upstream vessel and the frequency of the contractility of the adjacent vessel to be similar, due to the propagation of action potentials along lymphatic muscle cells adjacent to one another between lymphatic valves [50]. Similar contraction frequency between adjacent lymphatics has also been observed in vivo [46] and it has previously been suggested with experimental data that flow mediated dilation could assist in the coordination of contraction [28]. Thus, the ability to dynamically respond to flow, and the fact that the kinetics of this response is optimized to occur on time scales similar to the vessel's intrinsic contraction frequency, might provide a mechanism to enhance ejection fraction and optimize lymphatic pumping even in the presence of lymphatic contractile coordination via electrical coupling of lymphatic muscle cells. This is supported by Fig. 5a which shows that there is a significant increase in the ejection fraction when the externally applied flow is oscillating at a frequency close to the intrinsic frequency of contraction of the vessel. Similarly Fig. 5b shows a significant reduction in the fractional pump flow when the applied flow frequency is much lower than the intrinsic frequency of the lymphangion, due to the externally applied flow lowering the contractile frequency below its intrinsic value. Computational modeling that incorporates these dynamic responses, with values from actual experimental data such as that reported here, could shed light on the benefit of this flow mediated entrainment, possibly by preserving energy expenditure by lymphatic muscle cells and enhancing overall lymph transport.

The literature has also shown that the mechanosensitivity of lymphangions to shear stress is hampered in pathological states that have been related to impaired lymphatic transport, such as might occur in the case of metabolic syndrome [51,52] and as a result of aging [11,53]. The entrainment of lymphangions to shear stress may also be hampered during cases of lymphatic endothelial dysfunction that affects flow mediated dilation, thus leading to deficient pumping in the lymphatics or the unnecessary expenditure of energy by lymphatic muscle cells. This could provide a mechanism for reduced lymphatic pump flow during conditions of lymphatic dysfunction, possibly leading to impaired tissues-fluid homeostasis and other complications that may arise due to a compromised lymphatic system. Thus the physiologic consequences of impaired flow mediated dilation to lymphatic pump performance in the context of disease is an important area of future study.

## Conclusions

The lymphatic vessels are highly attuned to their local mechanical microenvironment as is reflected by the dependence of the shear sensitivity of the lymphatics on the diameter of the vessel. The shear dependent inhibition leads to an entrained contraction of the lymphangions to an oscillatory stimulus, thus pointing to a physical mechanism by which coordination of pumping between lymphangions might occur, leading to optimized lymph flow. The molecular mechanisms regulating this entrainment are not completely understood and future work should investigate the role of endothelial derived relaxation factors on this entrainment. Computational models can be improved with the information about the dependence of shear sensitivity on the diameter of the vessel, which can then be used to investigate how the coordination between lymphangions can affect the lymphatic pump function and how this might be compromised in disease.

## Methods

### *Experimental Setup*

The experiments were performed on thoracic ducts isolated from male Sprague-Dawley rats that weigh between 280-300 gm. All procedures on the rats were performed according to the relevant Institutional Animal Care and Use Committee (IACUC) guidelines at Georgia Tech (protocol number A14069). The thoracic duct was chosen since it has been observed that they show more sensitivity to shear stress variations than mesenteric lymphatic vessels [34]. The isolated vessels were cannulated in a vessel chamber and immersed in and perfused with physiological salt solution (PSS). The experimental setup consists of a commercially available vessel chamber from Living Systems Instrumentation, connected to a custom perfusion system that allows the independent control of transmural pressure ($P_{avg}$) and pressure gradient ($\Delta P$) along the vessel chamber through explicit model predictive control [54] as shown in Fig. 6. The whole setup was mounted on an inverted microscope, which was used to capture bright-field images of the vessel as it was exposed to the various flow conditions. Two micropipette tips mounted on the vessel chamber were used to cannulate the isolated thoracic duct. The size of the thoracic ducts ranged between 500 μm to 900 μm. Keeping this in mind, pipette tips that are approximately 450 μm in diameter were chosen. The vessel chamber was heated using electric heating pads and the temperature was maintained around 38°C using a thermocouple and a temperature controller. To ensure that the composition of the media did not change during the course of the experiments, fresh media was recirculated in the chamber using a peristaltic pump, running at a flow rate of 0.3 mL/min.

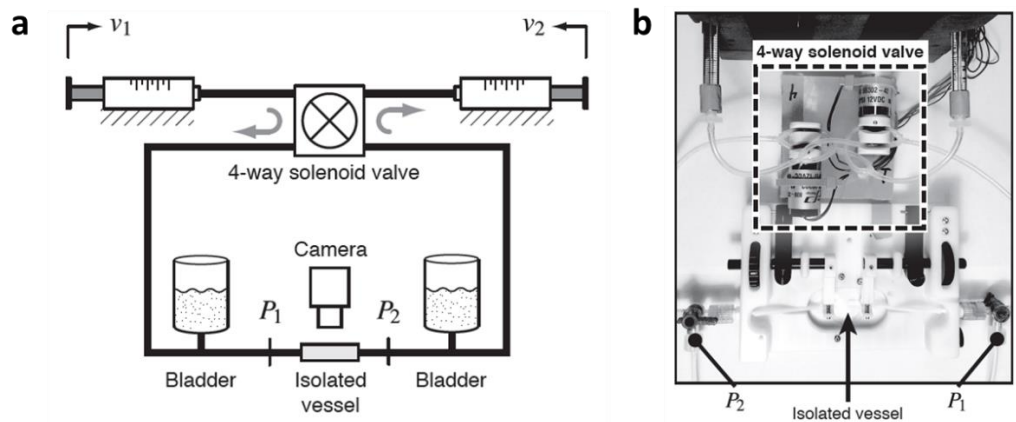

Fig. 6: The vessel perfusion system is shown highlighting the major components. a) The basic control scheme of the perfusion device is shown in the cartoon [54]. The solenoid valves ensure unidirectional flow through the chamber, from the inlet to the outlet. b) The top view shows the tubing connections from the device to the vessel chamber.

The mechanical conditions inside the vessel were controlled by the perfusion system that utilizes linear motors, gas-tight syringes and solenoid valves to control the pressures at the inlet and outlets of the vessel chamber, P1 and P2 respectively [54]. Analog pressure transducers were used to detect these pressures and control the device. The transmural pressure was calculated as the average of the inlet and outlet pressures, (P1+P2)/2. The pipette tips were resistance matched and the length of the tubes connecting each pipette to the external tubing were matched to ensure that the pressure drop in either pipette are the same. The matching ensures that there is no flow through the vessel when the pressure at the inlet and outlet of the vessel chamber are the same, or in other words, there is no offset in the relationship between the pressure gradient across the vessel and (P1-P2).

*Protocols*

Each vessel was first stepped through transmural pressures ($P_{avg}$) of 1, 3, 5 and 7 cmH$_2$O under zero pressure gradient (ΔP) over 12 minutes to pre-condition them. This was done to ensure that the vessel isolation and cannulation process did not damage the vessel and that reference contraction frequency for the vessel can be determined, which is useful in the subsequent steps. After pre-conditioning, the vessel was subjected to 5 minute and 15-minute ΔP ramps from 0-9 cmH$_2$O at a $P_{avg}$ of 3 cmH$_2$O. The vessels were then taken through a series of oscillatory flow waveforms of different frequencies and amplitudes. Each step consists of 5 minutes of zero ΔP, and then 5 minutes of oscillatory ΔP, at a $P_{avg}$ of 3 cmH$_2$O. This $P_{avg}$ was chosen as 3 cmH$_2$O since the contractile function of thoracic duct (pumping frequency and stroke volume) was found to be optimum at this pressure. At no point was the vessel subjected to a retrograde flow. The frequencies of the applied flow waveforms were 0.075 Hz, 0.2 Hz, and 0.35 Hz. These frequency values were chosen to correspond to frequencies much less than, similar to, and much greater than the intrinsic contraction frequency of about 0.3 Hz, so that an even distribution of data as a function of frequency is obtained during the data analysis. The amplitudes of the ΔP were chosen to be 4, 6 and 8 cmH$_2$O in order to correspond to "low", "medium" and "high" shear stresses as compared to the critical shear stress at which inhibition occurs (typically between 3-7 cmH$_2$O for the rat thoracic duct). This choice of the ΔP also ensured an even distribution of data as a function of shear stress during the data analysis.

*Data Analysis*

The data was acquired in the form of brightfield videos of the vessel acquired at 4x magnification. The videos were then processed frame by frame by a window-based thresholding algorithm that helped distinguish the vessel from the background as shown in Fig. 7a and 7b. Processing was also done at this step to remove any noise due to debris floating in the chamber. The diameter at each location along the vessel was then calculated by detecting the upper and lower boundaries of the vessel in the thresholded image. The diameters were detected over time to form the diameter tracings (Fig. 7b) that were then subsequently analyzed at each location along the vessel. Once the diameter data was obtained as a function of time and location, the tracings were filtered using an averaging low-pass filter to remove any noise due to the low resolution of the image.

The diameter tracings were then analyzed with Continuous Wavelet Transform (CWT) to accurately obtain the frequency distribution in the diameter tracings. The most commonly used tool in the analysis of frequency domain information is the Fourier Transform (FT), but it shows the magnitude of the frequency content of the data over a time window. When the spectral information is needed as a function of time, a modification to this method can be used which is called the "Short Time Fourier Transform" (STFT). In this technique, the spectral information is obtained over user-defined time windows within the data, where the windows overlap with each other by an amount defined by the user. Hence, it is apparent that in this method, the temporal resolution is limited by the amount of overlap, and the user-defined nature of the time window can change the amount of information captured. For example, a smaller time window can be used when the data is rich in higher frequencies and a larger window should be used when the frequency in the data is more concentrated in the low frequencies, so as not to miss out on any information. These problems can be avoided using Continuous Wavelet Transforms as it accounts for the varying time and frequency resolution without any user intervention, thus reducing human interference to the analysis. A caveat to this technique is that interpreting the data is not as straightforward as in Fourier transforms. In an FFT, the magnitudes simply represent the amplitude of the corresponding frequency component in the data. Wavelet transforms, however, utilize wavelets as the basis for the transform instead of cosines as in the case of FT. The use of these bases makes it harder to have a physical interpretation of the data as in the case of FT, but it provides higher temporal and frequency resolution, thus enabling the study of the spectrum of the signal with higher fidelity.

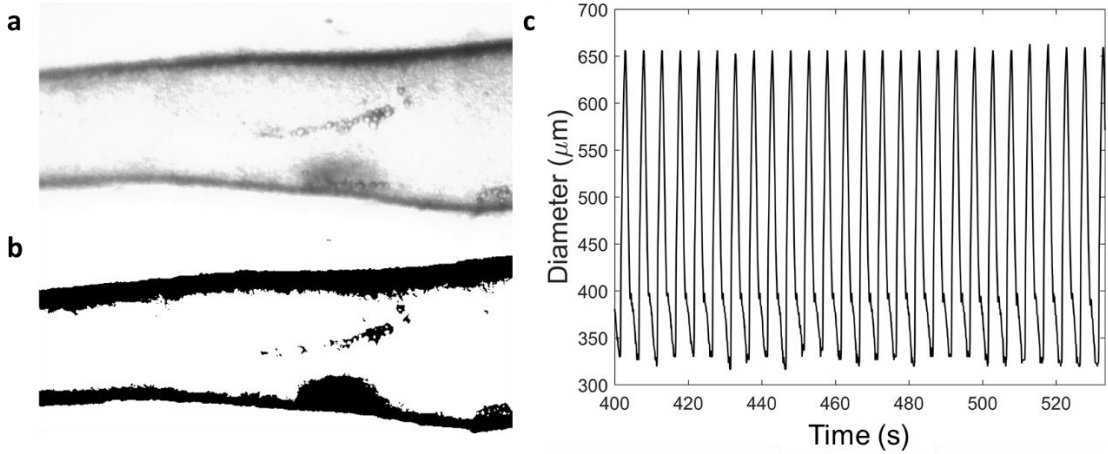

Fig. 7: The data analysis to obtain the diameter tracings from the brightfield images is shown. a) Brightfield images of the vessels are thresholded to get b) binary images of the vessel, isolated from the background. The binary images are used to detect the diameter of the vessel at different locations. v) Going through all the frames in the video, the diameter tracing vs time can be obtained for all the locations along the vessel.

### *Calculation of Wall Shear Stress*

The pressure difference across the vessel chamber is not a reliable estimate of the wall shear stress on the lumen of the vessel. There are minor losses associated with the sudden expansion of the pipette tip into a much larger vessel lumen. The pipette tip diameter is fixed, but the vessel diameter is variable, and hence the effective pressure drop across the vessel becomes a function of the vessel diameter. Hence the shear data for these experiments were obtained from the information of the location of the syringes as a function of time. Due to noise in the location data of the syringe positions, the location of the syringes were tracked in 3-second intervals to obtain an average syringe velocity over that time interval. After this time averaged syringe velocity ($V_{avg}$) was obtained, the flow rate, Q, was calculated from the knowledge of the inner radius, r of the syringe as

$$Q = \pi r^2 V_{avg}$$

From the diameter tracings of the vessel, an average diastolic diameter of the vessel, $D_{diast}$ was calculated. Using this diameter value, the representative wall shear stress was calculated as

$$\tau_{wall} = \frac{32 \mu Q}{\pi D_{diast}^3}$$

Where µ is the dynamic viscosity of water at 38°C which is around 6.78 mdyn-s/cm$^2$. Given the low Reynolds number (i.e. Re<<1) and low Womersley number (Wo<0.1) for these flows [55], and the fact that relatively straight vessel segments without valves are chosen, the Poisuielle flow approximation could be used to get an appropriate estimate of the applied WSS.

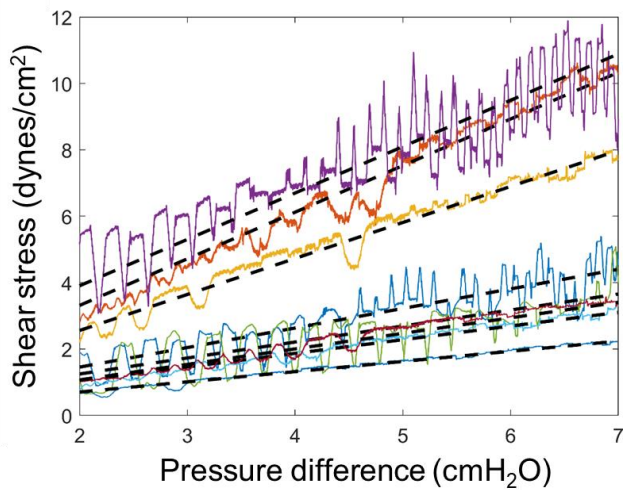

Fig. 8: For the ramp shear stress experiments the wall shear stress is plotted as a function of the pressure difference and straight lines are fitted to them. The fitted pressure to shear conversion graphs are used to convert the oscillatory pressure to oscillatory shear stress.

Once the shear stress data was obtained for the ramps, it was used to convert the ΔP to WSS to be used in the conditions where an oscillatory shear stress was applied. This conversion is required since the syringe position is averaged over 3 seconds to be reliable, a time window that is too long to capture the shear stress variations in high-frequency oscillatory flow conditions. This estimation was done by obtaining the ramp shear stress as a function of ΔP and then plotting straight lines to them, as shown in Fig. 8. The pressure to WSS relationship was obtained for each vessel individually. The slope and offset of the line fits were used to calculate the OWSS from the oscillatory ΔP for each vessel. The advantage of calculating the OWSS in this way is that there is no reliance on the ΔP across the system for calculating the WSS, hence eliminating the variability introduced in the signal due to changes in the vessel diameter.